\documentclass[runningheads]{./llncs}
\usepackage[noframe]{showframe}

\usepackage{amsmath}
\usepackage{gnuplottex}
\usepackage{graphicx}
\usepackage{subcaption}
\usepackage{xparse}
\usepackage[hidelinks]{hyperref}

\usepackage{url}
\usepackage{booktabs}
\usepackage{tabularx}

\begin{document}

\title{zeek-osquery: Host-Network Correlation for Advanced Monitoring and Intrusion Detection}
\titlerunning{zeek-osquery}
%

\author{
	Steffen Haas\inst{1} \and
	Robin Sommer\inst{2} \and
	Mathias Fischer\inst{1}
}
\authorrunning{S. Haas et al.}
%
\institute{
	Universit\"at Hamburg, Hamburg, Germany\\
	\email{\{haas,mfischer\}@informatik.uni-hamburg.de}
	\and
	Corelight, Inc., San Francisco, CA, USA\\
	\email{robin@corelight.com}
}

\maketitle

\begin{abstract}
Intrusion Detection Systems (IDSs) can analyze network traffic for signs of attacks and intrusions. However, encrypted communication limits their visibility and sophisticated attackers additionally try to evade their detection.
To overcome these limitations, we extend the scope of Network IDSs (NIDSs) with additional data from the hosts. For that, we propose the integrated open-source \emph{zeek-osquery} platform that combines the Zeek IDS with the osquery host monitor. Our platform can collect, process, and correlate host and network data at large scale, e.g., to attribute network flows to processes and users. The platform can be flexibly extended with own detection scripts using already correlated, but also additional and dynamically retrieved host data. A distributed deployment enables it to scale with an arbitrary number of osquery hosts.
Our evaluation results indicate that a single Zeek instance can manage more than 870 osquery hosts and can attribute more than 96\% of TCP connections to host-side applications and users in real-time.
\keywords{intrusion detection \and network monitoring \and host-network correlation \and zeek \and osquery.}
\end{abstract}

\section{Introduction}
\label{sec:intro}

Computer networks need a second line of defense against cyber-attacks, in which network devices and connected systems are monitored to detect signs of intrusions. NIDSs can fill this gap and allow to collect extensive information about a monitored network. They can detect ongoing attacks and compromised hosts. 
However, the (definitely positive) ongoing trend towards secure and encrypted communication, turns an IDS partially blind. It cannot analyze the encrypted data anymore and thus might miss signs of intrusions.
Furthermore, the complete reconstruction of sophisticated attacks, e.g., Advanced Persistent Threats (APTs)~\cite{friedberg2015combating}, is almost impossible based on network data only. Especially the detection of multi-step attacks across multiple hosts~\cite{shin2019unsupervised} that requires to identify related or similar flows, e.g., the Command and Control channel after a Trojan download and the following lateral movement in the network, can only be correlated with uncertainty without any insights into host data~\cite{wilkens2019towards}.

Host context on network flows can improve the accuracy of NIDSs~\cite{snapp1991dids,almgren2001application,dreger2005enhancing}. The combination of host and network monitoring leads to an increased visibility on attacks and requires to jointly analyze host and network data. However, while an NIDS can protect a complete network, a Host IDS (HIDS) has to run on every device in the network. Moreover, this induces a high correlation load when correlating host with network data. Furthermore, to be applicable for intrusion detection, such a system must be able to process the data of the monitored hosts close to real-time and thus need to scale with the number of HIDSs.
Security Information and Event Management (SIEM) systems have been designed for this task~\cite{bhatt2014operational}. They are centralized systems that usually read in log files collected from NIDSs and HIDSs and perform high-level correlation and aggregations across them. 
In contrast to their coarse-grained data that lacks detailed network data, our fine-grained correlation system makes use of causal relations in the data for the purpose of real-time intrusion detection in the network.

The main contribution of this paper is the novel open-source platform \emph{zeek-osquery}\footnote{\url{https://github.com/zeek/zeek-osquery}} for the scalable and joint monitoring of networks and their hosts. For that, we combine the network monitor and IDS Zeek~\cite{paxson1999bro} (formerly known as Bro) with the host monitor osquery~\cite{Osquery}. 
Our solution correlates data from the Operating System (OS) level with network information in real-time. 
Furthermore, it allows to dynamically select from a great variety of OS data available for processing. This way, we provide the fundamentals for a new class of detection algorithms that operate on a much broader visibility. We extend the context of network flows, for example to attribute them to the originating processes and the users that started them. 
Zeek-osquery can be flexibly adapted to different detection scenarios, as osquery-hosts are directly managed from Zeek scripts and all data processing can be implemented in Zeek. Examples are the detection of executed files downloaded from the Internet, the detection of lateral movement of attackers via SSH hopping~\cite{wang2018research}, or to provide Zeek with Kernel-TLS keys obtained at hosts for the decryption and inspection of network traffic.

We extensively evaluated zeek-osquery in a small-scale real-world deployment and conducted additional experiments to investigate its scalability with an increasing number of osquery hosts.
Our evaluation results indicate that we can attribute more than 96\% of all TCP connections to their originating processes and the responsible users on monitored hosts, contrary to less than 0.1\% attributed connections when using Zeek alone. Moreover, our system seems to scale with an increasing number of osquery hosts, enabling one Zeek instance to handle more than 870 osquery hosts in our evaluation setting. For larger deployments, we also propose a distributed setup with multiple Zeek instances that enable zeek-osquery to scale to arbitrarily large networks.

The remainder of this paper is structured as follows: Section~\ref{sec:related_work} presents related work. Section~\ref{sec:refining} highlights the concept of host-network correlation and shows how to link host and network data for network attribution. Section~\ref{sec:system} introduces our open-source platform \emph{zeek-osquery} to monitor both hosts and network and to correlate monitoring data in real-time. The evaluation of zeek-osquery in Section~\ref{sec:eval} is done in a real-world deployment for insights and in a stress test to highlight the scalability of our solution. We conclude with Section~\ref{sec:conclusion}.

\section{Related work}
\label{sec:related_work}

We identified four groups of related work that is relevant for our work in the following: (1) Collaborative IDS (CIDS), (2) host context for network intrusion detection, (3) host activity to describe communication behavior, and (4) SIEM systems that can correlate logs from several sources.

\textit{CIDSs} consist of several host or network IDSs that collect, exchange, and analyze data to create a holistic view of a network.
However, these CIDS either do not scale, are built for a very specific purpose, e.g., to detect worm outbreaks~\cite{cai2005collaborative}, or offer a poor detection accuracy~\cite{Vasilomanolakis2015}.

From the perspective of a NIDS, the inclusion of \textit{environmental context} can increase the detection accuracy~\cite{sommer2003enhancing}. 
An early work from Snapp et al.~\cite{snapp1991dids} combines a NIDS with remote login events from hosts to identify login chains across multiple hosts for the same user.
More recently, the context directly comes from the communicating application.
Almgren et al.~\cite{almgren2001application} instruct a NIDS to verify that the application processed the network messages correctly and to retrieve decrypted message payloads.
Dreger et al.~\cite{dreger2005enhancing} compare how the NIDS and the application decode network messages. This allows to detect attackers that obfuscate their traffic for IDS evasion.
However, these approaches require modifications to all applications and do not systematically correlate host and network data to increase the network visibility in general.

To gain more insight into \textit{host activities} related to network communication, dependency graphs with processes, files, and sockets are utilized to link host activities~\cite{king2003backtracking}. The objects in the dependency graph are usually created from OS audit data~\cite{bates2015trustworthy}.
For example, Ma et al.~\cite{ma2016protracer} taint objects in these graphs to propagate provenance in audit logs. This allows to enrich network-related activities on a single host, but it is not incorporated into network intrusion detection.
For the usage of dependency graphs in intrusion detection~\cite{liu2019host}, 
King et al.~\cite{king2005enriching} extend these graphs across hosts to argue for causal relationships between attack steps.
Sun et al.~\cite{Sun2016CNS} incorporate IDS alerts into dependency graphs linked across hosts to find the most likely attack path through the network.
However, to the best of our knowledge, no approach systematically leverages host context similar to dependency graphs for network intrusion detection in real-time.

\textit{SIEM} systems~\cite{bhatt2014operational} store logs from various sources, including host monitors, host applications, and network monitors in a central storage. They correlate host and network data to detect and investigate security incidents.
However, SIEM systems fail to detect sophisticated attacks, because they miss detailed network data and their capabilities for analysis are usually limited to data aggregation, thresholding, and pattern detection. In contrast, our work provides additional host context specifically for particular network flows. It is used flexibly in Zeek scripts for network intrusion detection in real-time. 
Analysis results can be written to log files, be processed by SIEM systems, or induce an alert.

\section{Refining the network visibility}
\label{sec:refining}

To increase the accuracy of intrusion detection, we highlight network-related host activity that links between network and host monitoring.
More generally, we refine the network visibility by incorporating host context into network monitoring. For all network traffic, we identify related properties that come from the hosts, e.g., user or process information. In particular, we explain how to attribute traffic in the network to applications and users on the respective hosts.

\subsubsection{Network-related activities on hosts}
\label{subsec:refining_events}

Throughout this paper, we use the term \textit{flow} to describe the communication between two hosts with a 5-tuple of IP address and port of the hosts, and the protocol.
We denote the flow initiator as \textit{originator} and the other host as \textit{responder}.
\textit{Processes}, 
identified by a process ID (\textit{pid}), use sockets to abstract flows. A socket is identified by a unique socket ID, i.e., the combination of a file descriptor (\textit{fd}) and pid, and additionally includes the attributes of the respective flow 5-tuple.
Processes and sockets are retrieved by monitoring the system calls (\textit{syscalls}) to the kernel, e.g., as provided by the kernel audit in Linux. Several syscalls exist for the interaction of processes with the kernel, including \texttt{execve} to spawn processes as well as \texttt{bind} and \texttt{connect} to establish incoming and outgoing flows, respectively.
Another way to retrieve processes and sockets is to probe the current kernel status of Windows, Mac, and in particular the \texttt{procfs} in Linux. Kernel status data holds all attributes about current processes and sockets.

The data from both \textit{kernel audit} and \textit{kernel status} come along with different properties.
While the audit allows for an asynchronous pushing of new processes and sockets, the status has to be frequently probed and compared with the previous one to detect any changes. 
Despite this probing overhead, it outperforms the audit variant when looking at the available attributes (data \textit{soundness}). This is because audit monitors syscalls and consequently can only record actual parameters of these calls. For example, in case of a \texttt{connect}, only the destination IP address and port are part of the call. The local IP address and port remain unknown, even when combining several socket-related syscalls.
However, relying on the status variant alone is error prone regarding retrieving all objects (data \textit{completeness}), as a short-living process or socket might start and end between two status probes. Consequently, such processes or sockets would be missed.
Therefore, a combination of both audit and status variants is elaborated next to achieve both full data soundness and completeness.

\subsubsection{Attributing network flows}
\label{subsec:refining_heuristic}

\begin{figure}[t]
	\centering
	\includegraphics[trim={0 0 0 0},clip,width=1.0\linewidth]{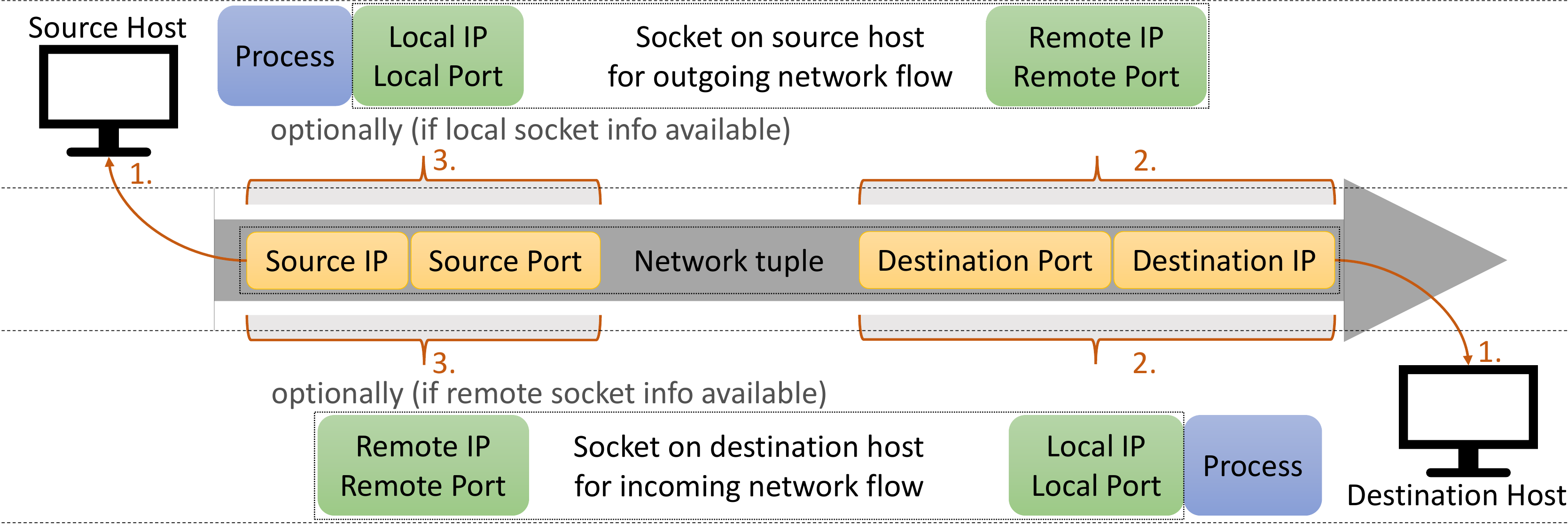}
	\caption{Attributing a network flow to a process on source and destination host by matching the network tuple to the socket information on each of the hosts.}
	\label{fig:correlation_heuristic}
	\vspace*{-12pt}
\end{figure}

Both the originator and the responder of a network flow can be attributed by identifying the respective socket on either host, where applicable. For simplicity, we assume that the socket identification is sufficient, as this is the missing link to other host activity like processes and files~\cite{bates2015trustworthy}.
The identification of the respective socket for a network flow requires a match on the flow 5-tuple. On the originator, this is a socket representing an outgoing flow, and an incoming flow on the responder. This is relevant as the 5-tuple for sockets reflects the IP address and port of the local and remote hosts, while it reflects originator and responder host for network flow. Thus, the incoming socket actually requires to match the inverse 5-tuple as the destination in the network flow is required to match the local host on the responder. Figure~\ref{fig:correlation_heuristic} illustrates the originator with its outgoing socket (top left), the responder with its incoming socket (bottom right), and the network flow with its full 5-tuple (middle).

In contrast to kernel status data, the incomplete data soundness from kernel audit results in unavailable socket attributes to match the source of a network flow. To account for that, we correlate sockets and network flows for attribution as following in three steps:
\begin{enumerate}
	\item[(1)] Identify originator and responder hosts by the IP addresses in the network flow. This requires a maintained list of IP addresses and hosts in the network.
	\item[(2)] On the originator and responder, identify the socket(s) for which the flow destination equals the remote or local socket info, respectively.
	\item[(3)] Also require the flow source to equal local or remote socket info, respectively.
\end{enumerate}
The correlation is unambiguous when the socket attributes for Step 3 are available. Otherwise, the correlation might be \emph{vague}. In case of two hosts with a \texttt{connect} syscall to the same destination IP address and port, our correlation is still unambiguous because of Step 1. However, it is vague for the same host with multiple flows to the same remote IP and port (from different source ports).
Ideally, the correlation outcome is exactly one socket for the originator and the responder. However, in case of vague correlation we list all candidate hosts, processes, and users that might be responsible for the flow.

\subsubsection{Validity of activities}
\label{subsec:refining_validity}

Processes and sockets that already terminated some time ago must not be considered for attribution of a currently ongoing network flow. For that, we hold \textit{state} about host activities and remove them from the state when the respective process terminated or socket is closed. Thus, with maintaining the host activities according to their \textit{validity} in our state, we aim to follow the life-cycle of processes and sockets from their creation to termination.

Data from kernel status is easy to incorporate into the state, as the status reflects a current snapshot and comparing it to the previous snapshot allows to explicitly identify new or removed processes and sockets, respectively. For the kernel audit, the life-cycle for processes and sockets can be followed by syscalls like \texttt{execve}, \texttt{socket}, or \texttt{close}. This works as long as the process decides on its own to terminate or to close a socket. But there will not be such an audit event when the process crashes or the TCP connection breaks. To prevent our state to be polluted in such cases, we regularly perform a \textit{verification} of it. We do so by probing the host if all the processes (pid) and sockets (pid, fd) in our state are still present by the current kernel status.

\section{Monitoring and event correlation with zeek-osquery}
\label{sec:system}

In the following, we will describe zeek-osquery, a system for the collection, analysis, and correlation of host and network data that follows the approach from Section~\ref{sec:refining}.
After introducing the existing monitoring tools Zeek and osquery, we explain how our system \emph{zeek-osquery} combines these two tools for the purpose of advanced monitoring and event correlation.

\subsection{Monitoring tools and overview}
\label{subsec:overview}

The network monitoring and intrusion detection is performed by \emph{Zeek}~\cite{paxson1999bro} (formerly known as Bro). It captures and parses network traffic in its core and provides a powerful scripting language to further analyze the traffic with custom scripts.
The Zeek publish-subscribe (pub-sub) library Broker\footnote{\url{https://docs.zeek.org/projects/broker}} dispatches events internally even among remote machines.
We have enhanced Zeek to retrieve host events from osquery hosts and to natively process them analogous to network events in Zeek. Based on this, Zeek exposes itself as a platform with the ability to correlate various information from both host and network events.

\emph{Osquery}~\cite{Osquery} is an OS instrumentation framework. It provides an SQL-like interface to query the OS as a relational database, including from kernel audit and status. SQL tables represent abstract concepts such as running processes, open network flows, browser plugins, or file hashes.
When running in background as host sensor, osquery regularly executes SQL queries as defined in its schedule. 
We extended osquery to communicate with Zeek via Broker to retrieve SQL requests from Zeek and report matching host events. The flexible pub-sub communication enables osquery hosts to join and leave the Broker overlay at any time. The seamless integration of host events from osquery into the Zeek processing pipeline allows for efficient processing of host events and their correlation with network events directly in Zeek scripts.

In zeek-osquery, the hosts are continuously monitored by osquery, which send host events to Zeek via the Broker overlay that connects osquery hosts and Zeek among each other. Added or removed entries in an osquery table are retrieved as a continuous stream of host events. Alternatively, osquery tables can be also queried on demand to retrieve a snapshot of an osquery table, e.g., to gather all running processes at a certain time. This allows for an interactive analysis of hosts, e.g., to investigate a specific security issue.
Zeek acts as a correlation platform that analyzes and correlates both, the network events for the traffic captured by Zeek and the host events retrieved from all osquery hosts. For that, we extended Zeek by a new Zeek framework to define custom queries and result handling. For large deployments, our Zeek correlation platform can be set up in a distributed manner with multiple, communicating Zeek instances.

\begin{figure}[t]
	\centering
	\includegraphics[trim=0 0 0 0, clip, width=0.8\linewidth]{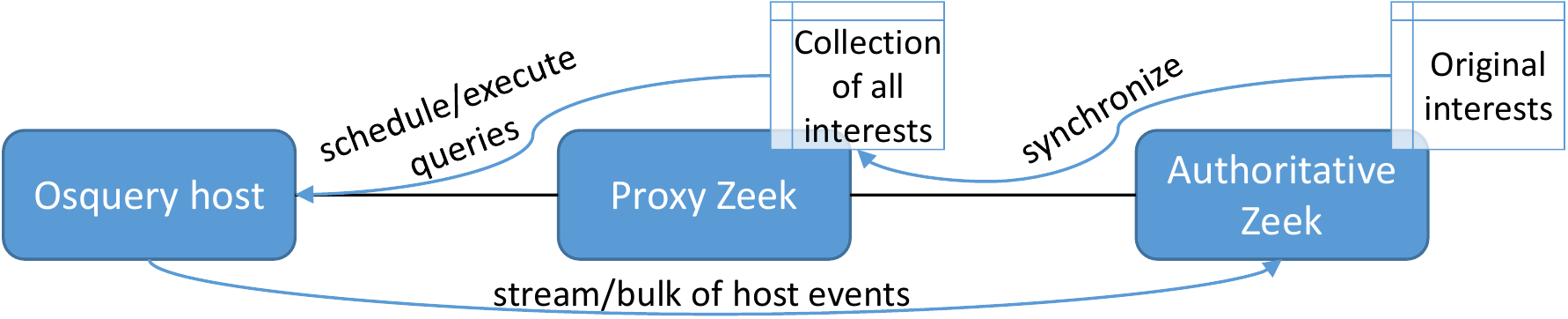}
	\caption{Communication among osquery and Zeek.}
	\label{fig:zeek_proxy}
	\vspace*{-12pt}
\end{figure}

\subsection{System architecture}

We implemented a novel Zeek \emph{framework}, i.e., a collection of Zeek scripts, to control osquery hosts. 
This way, Zeek is capable of: (1) requesting complete results to a one-time query immediately and (2) scheduling queries that are regularly executed.
We address specific osquery hosts or groups of them in the pub-sub overlay by distinct topic names, labeled as \texttt{groups} throughout this paper. Apart from some default \emph{groups}, custom ones can either be pre-configured at the hosts or dynamically controlled by Zeek.
It uses the group labels to control the SQL queries for specific selections of osquery hosts. An \texttt{interest} denotes the binding of a query to a \emph{group}. It contains additional information, e.g., whether the query is executed regularly or just once and how to send the results back to Zeek.
This way, Zeek can publish an \emph{interest} over Broker to osquery hosts in a particular \emph{group}, e.g., for logged in users on all monitored servers.

When an \emph{interest} is published, the currently connected osquery hosts will receive it, but others that join the overlay later missed previous but valid \emph{interests}. Thus, they would not execute the scheduled queries. For that, Zeek instances can take over the following roles as illustrated in Figure~\ref{fig:zeek_proxy}:
\begin{itemize}
	\item An \texttt{Authoritative} Zeek is the origin of an \emph{interest}. This role defines queries and retrieves query results, i.e., host events, from the osquery hosts.
	
	\item A \texttt{Proxy} Zeek collects and holds all interests from \texttt{Authoritative} Zeek instances. It forwards only applicable queries to its directly connected osquery hosts. 
	The \texttt{Proxy} Zeek maintains the query schedule and \emph{group} assignments of hosts, both when they join and later when their interests change.
\end{itemize}

Our overlay design with respect to the Zeek roles, enables a scalable deployment with multiple Zeek instances that run in a distributed fashion for load balancing and availability reasons. Distributing the load among multiple Zeek instances can be achieved in three ways:
\begin{itemize}
	\item Resource intensive correlation tasks can run exclusively on an additional \emph{Authoritative} Zeek instance. For that, another Zeek joins the overlay and publishes \emph{interests} for events that are required for detection. The resources of this instance are completely available to the detection and all other instances can continue using their resources to perform their tasks.

	\item If large amounts of osquery hosts would overwhelm a single \emph{Authoritative} Zeek instance, the osquery hosts are organized in \emph{groups}, with one out of multiple Zeek instances being responsible for one \emph{group}. All Zeek instances are then interested in the same query that they publish to a specific \emph{group}.
	
	\item To reduce the load on a single \emph{Proxy} Zeek instance, multiple instances can be deployed. Then, each one needs to handle a lower number of directly connected hosts that still receive the same \emph{interests} as before.
\end{itemize}

After an osquery host joined the Broker overlay, it is controlled by its \emph{Proxy} Zeek and now accepts and executes any forwarded \emph{interest}.
However, note that \emph{interests} originally come from \emph{Authoritative} Zeeks and query results should also be routed back to them via the pub-sub overlay. For that, the originating \emph{Authoritative} Zeek by default sets the response topic to its own topic when publishing \emph{interests}.
If the same \emph{interest} query originates from multiple Zeeks, we suggest to choose the same response topic across Zeek instances for the same \emph{interest}. This way, \emph{interests} can be consolidated on osquery hosts and the query results are sent efficiently over Broker to multiple \emph{Authoritative} Zeek instances.

\subsection{Event correlation for network attribution}
\label{subsec:zeek_osquery_correlation}

\begin{figure}[t]
	\centering
	\includegraphics[trim={0 0 0 0},clip,width=1.0\linewidth]{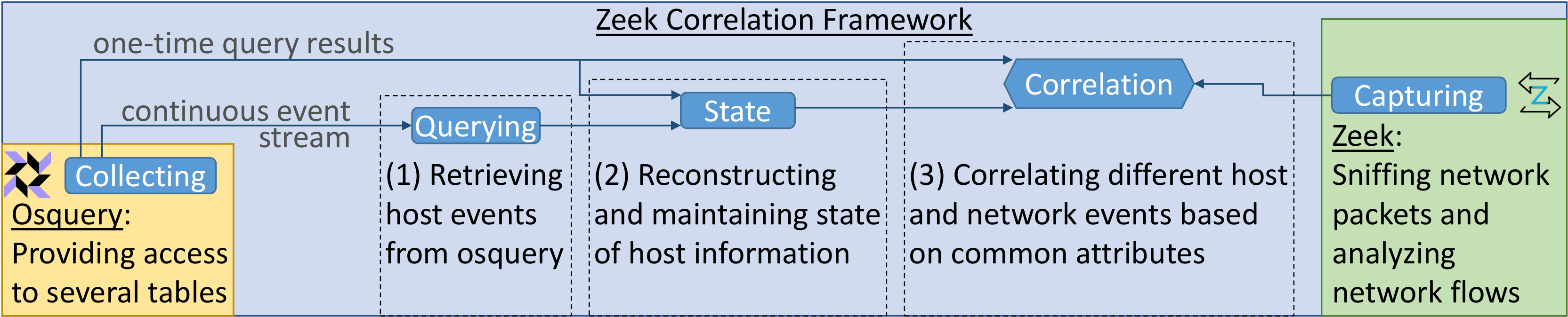}
	\caption{Architecture of the processing pipeline.}
	\label{fig:correlation_architecture}
	\vspace*{-12px}
\end{figure}

We developed a processing pipeline as part of our Zeek framework to process and correlate host and network events. We implemented it in the Zeek-typical event-based fashion to allow custom scripts to reuse events emitted by our pipeline for further analysis.
The three different stages in the processing pipeline are illustrated in Figure~\ref{fig:correlation_architecture}, follow the concept as described in Section~\ref{sec:refining}, and are detailed in the remainder of this section.

In the \texttt{Querying} stage \emph{interests} are defined, i.e., SQL queries, that are sent to and scheduled on osquery hosts. Events on this stage are a continuous stream of raw host events that directly come from osquery. Incoming events reflect updates of an osquery table, e.g., processes or sockets.

The \texttt{State} stage assembles raw host events from osquery to a state in real-time and reflects the current host status, e.g., a process is added upon creation and removed upon termination.
Tables in osquery based on Linux kernel audit (cf. Section~\ref{sec:refining}) report only new processes and sockets, so we verify the state periodically against the kernel status utilizing one-time queries.
The same mechanism is used to retrieve a snapshot for initial state before continuously updating it with audit events.
The state consolidates raw host events from different tables in case they describe the same class of data, e.g., the socket state is built based on the osquery tables \textit{socket\_events}, \textit{listening\_ports}, and \textit{process\_open\_sockets}. While state of Windows and Mac hosts is reconstructed solely from tables reflecting kernel status, state of Linux hosts is additionally based on kernel audit and therefore more accurate because of data completeness (cf. Section~\ref{subsec:refining_events}).

In the \texttt{Correlation} stage, the data base for correlations encompasses the triggering event and any state that is available both natively in Zeek and about osquery hosts.
As an example, we implemented the attribution of network flows by linking them with the respective application and user in real-time. To demonstrate the effect, we extended the statistics about every network flow in Zeek (\texttt{conn.log}) to additionally list the respective host, application, and user.

\subsection{Examples for Scenario Detection}
\label{subsec:zeek_osquery_scenarios}

Building on the network attribution (cf. Section~\ref{subsec:zeek_osquery_correlation}), we implemented the detection of three particular scenarios using our processing pipeline in Zeek scripts.

\paragraph{Execution of mail attachments}
Once an email campaign with malware is recognized and its mail recipients identified, it must be reconstructed who of them executed the attachment and on which machines.
Zeek-osquery tracks the execution of mail attachments as follows: First, Zeek notices a file download from the Internet as mail attachment and remembers the file hash associated with the attachment. Second, in parallel Zeek interactively requests the file hash of unknown binaries upon process creations. A match among the download hashes and the binary hashes reveals the execution of an Internet download.

\paragraph{Stepping Stone detection}
Attackers often hide their identify by using an infected machine in the network as proxy to reach the actual target. An example for such an attack is SSH chaining and zeek-osquery detects it as follows: First, Zeek identifies all hosts that have both an incoming and outgoing SSH connection, i.e., which is an indication for a SSH proxy. To verify the relation between incoming and outgoing connection, Zeek interactively requests the \emph{pid}s of all children under the process with the incoming connection. If this list contains the \emph{pid} of the process with the outgoing connection, a stepping stone is detected.

\paragraph{TLS decryption}
TLS proxies often actively break the end-to-end encryption for traffic analysis. With zeek-osquery, we provide the ability to selectively request cryptographic material from hosts such that Zeek can passively decrypt the traffic. As proof-of-concept, we extended osquery to capture the respective system call used in Kernel-TLS (KTLS), when the application forwards the keys to delegate the symmetric de- and encryption after the TLS handshake to the kernel. Osquery then provides a tables with the obtained keys to Zeek.

\begin{table}[t]
	\centering
	\caption{Characteristics of the real-world dataset.}\label{tab:dataset}
	\begin{tabularx}{\linewidth}{X|X|X|X|X|X|X|X}
		\multicolumn{4}{c|}{Network flows} & \multicolumn{4}{c}{Host events for state} \\
		Total   & TCP     & UDP    & ICMP            & Process   & Socket  & User   & Interface \\\midrule
		344,366 & 273,241 & 70,929 & 196             & 2,793,406 & 776,910 & 51,719 & 7,919 \\
	\end{tabularx}
	\vspace*{-12px}
\end{table}

\section{Evaluation}
\label{sec:eval}

In this section, we evaluate zeek-osquery. First, Section~\ref{subsec:real_eval} gives insights into a real-world deployment. Afterwards, we stress-test the system with more hosts and host events in Section~\ref{subsec:scale_eval} to evaluate its scalability properties.

\subsection{Real-world evaluation}
\label{subsec:real_eval}

We deployed zeek-osquery to monitor eleven office machines of a working group in the computer science department on a university campus for three working days. The machines were running different Linux distributions, including Ubuntu, Linux Mint, Fedora, and Arch Linux. 
To monitor their network traffic and to correlate it with host events, we tunnel the traffic through a VPN on the campus site that is monitored by a Zeek instance. Note, that the correlation is performed in real-time during this experiment.

The data processed by zeek-osquery in this setup is characterized in Table~\ref{tab:dataset}. It reports characteristics of the flows that our zeek-osquery ideally correlated with host data.
Furthermore, the table reports the number of received host events. On average, each of the eleven machines was monitored for $6$ hours and $21$ minutes per day. An individual host on average reported $20.22$ process, $5.63$ socket, and $0.37$ user events per minute that go into state (cf. Section~\ref{subsec:zeek_osquery_correlation}).
Note, that also the initial state that is retrieved from hosts upon their (re-)connects goes into the average event rate.
The following provides results and experiences on how zeek-osquery enhances the visibility and accuracy of monitoring.

\subsubsection{Attribution}
\label{subsubsec:real_attribution}

Table~\ref{tab:attribution_rate} shows the success rate of our attribution for the 344,366 network flows in our dataset. For $96.05\%$ of TCP connections and $86.61\%$ of all flows we identify the responsible processes and users. False negatives are caused when: 
First, the host data is not retrieved in time for real-time correlation with short-lived flows. 
Second, applications like \texttt{Skype} use Stateless IP/ICMP Translation (SIIT) to embed the actual IPv4 destination address into an IPv6 address. But as the host sends an IPv4 message, this causes a mismatch between IPv4 (in network flow) and IPv6 addresses (in host events). 
Third, remote hosts continue a flow although the monitored host already left the VPN and a new host joined the VPN reusing the same IP address. These packets cannot be attributed to a process on the new host.
The attribution rate for UDP flows is only about 50\% because Zeek retrieves host events about UDP sockets from the audit status only (cf. Section~\ref{sec:refining}), which is provided at discrete time slots only. Thus, short-living sockets might be missed out. This holds especially true for DNS requests that are responsible for $89\%$ of the UDP flows.

\begin{table}[t]
	\centering
	\caption{Attributing of network flows.}\label{tab:attribution}
	\begin{subtable}{.4\textwidth}
		\centering
		\caption{Attribution rate}\label{tab:attribution_rate}
		\begin{tabularx}{0.8\linewidth}{X|X|X}
			All & UDP & TCP \\\midrule
			86.61\% & 50.43\% & 96.05\% 
		\end{tabularx}
		\vspace*{10px}
	\end{subtable}%
	\begin{subtable}{.6\textwidth}
		\centering
		\caption{Attribution uniqueness}\label{tab:attribution_uniqueness}
		\begin{tabularx}{0.9\linewidth}{r|X|X|X}
			& Host    & Process & User            \\\midrule
			Unique attributions & 100\% & 88.53\% & 98.14\%         \\
			Average candidates  & 1.00    & 1.17    & 1.02
		\end{tabularx}
	\end{subtable}
	\vspace*{-12px}
\end{table}

We further evaluate the attributed flows with respect to a unique host, process, and user. In Table \ref{tab:attribution_uniqueness}, we count the number of flows that have been attributed to a single entity and furthermore calculate the average number of candidate entries per attribution. Apart from the vague correlation (cf. Section~\ref{subsec:refining_heuristic}), a fast re-usage on hosts of the same process ID or socket, i.e., file descriptor, can be a reason for multiple attribution candidates.
The effects of the vague correlation become visible, especially for DNS flows. Usually applications use the DNS server defined by the OS and therefore many processes establish flows to the same server and port combinations. If we skip attributing flows to the DNS servers, the unique attribution of processes increases from 88.53\% to 93.15\%.
Although a single user was logged in on the monitored machines, in some cases the user attribution overlaps with a system account, e.g., in case of parallel DNS requests by the system and an user application.

\subsubsection{Network applications}
\label{subsubsec:real_netapp}

To identify communicating applications, the state-of-the-art is to inspect network packets for application specific indicators like the HTTP user agent. Zeek already analyses such indicators and derives the respective application, where applicable.
If Zeek itself cannot derive the application from the network packets, zeek-osquery can still verify the application via the correlation with host data. Table~\ref{tab:netapp} lists the top 10 network applications ranked by their number of attributed flows.
We observed two outcomes when comparing both methods for identifying communicating applications: 
First, zeek-osquery is able to attribute significantly more flows compared to Zeek, i.e., 298255 (86.61\%) compared to 212 (0.06\%). Specifically for the Firefox browser, zeek-osquery was able to attribute flows 2971 times more often than Zeek. 
Second, zeek-osquery is able to identify applications that were not identified by Zeek. This includes user applications such as \texttt{Syncthing}, \texttt{Seafile}, and \texttt{Skype}, but also system-related components such as the network time synchronization daemon \texttt{NTPD} and the Dynamic Host Configuration Protocol (DHCP) client \texttt{dhclient}.

\begin{table}[t]
	\centering
	\caption{Top 10 attributed applications among all network flows.}\label{tab:netapp}
	\begin{tabular}{c|rr|rr}
		& \multicolumn{2}{c|}{Zeek}  & \multicolumn{2}{c}{zeek-osquery} \\
		Rank & \multicolumn{1}{l}{Attributed flows:} & \multicolumn{1}{r|}{0.06\%}  & \multicolumn{1}{l}{Attributed flows:} & \multicolumn{1}{r}{86.61\%} \\\hline
		1  & Chrome               & (0.01\%) 
		& Firefox              & (23.17\%) \\ 
		2  & Firefox              & (0.01\%) 
		& Thunderbird          & (12.30\%) \\ %
		3  & Spotify              & (0.01\%) 
		& Spotify              & (6.11\%) \\ %
		4  & Thunderbird          & (0.01\%) 
		& Opera                & (5.41\%) \\ %
		5  & Debian APT-HTTP      & ($<$0.01\%) 
		& Syncthing            & (5.39\%) \\ %
		6  & libdnf               & ($<$0.01\%) 
		& Chromium             & (4.55\%) \\ %
		7  & Wget                 & ($<$0.01\%) 
		& Skype                & (3.87\%) \\ %
		8  & $<$unknown browser$>$& ($<$0.01\%) 
		& Seafile              & (3.80\%) \\ %
		9  & OpenSSH              & ($<$0.01\%) 
		& Chrome               & (3.56\%) \\ %
		10 & gvfs                 & ($<$0.01\%) 
		& qutebrowser          & (3.33\%) \\ %
		
		\hline
		
		Total & \multicolumn{2}{c|}{33 applications} & \multicolumn{2}{c}{88 applications}
	\end{tabular}
	\vspace*{-12px}
\end{table}

However, we have also seen limitations of zeek-osquery, especially when a process launches another application that immediately starts a network flow. Because the flow could have happened before or after the parent process with the same pid transferred control with the \texttt{execve} syscall to the new child, both parent and child application are candidates for the attribution. In our experiment, applications that are known to never communicate directly are candidates for $0.18\%$ of attributed flows.
Also, some monitored hosts were running NATed Virtual Machines (VMs) with a Windows guest system. While osquery runs on the Linux hypervisor host only, it attributes any traffic of the VM to the virtualization application ($2.29\%$). However, Zeek might still identify the Windows application inside the VM based on identifiers in the network packets ($0.01\%$).

Zeek-osquery significantly increases the identification rate of communicating applications. This enables to enforce the use of allowed applications and assists threat hunters in detecting malware that covers its communication in well-known protocols, e.g., HTTPS, that is usually allowed to pass the firewall.

\subsection{Zeek performance analysis}
\label{subsec:scale_eval}

In this section, we provide evaluation results to assess the scalability and efficiency of zeek-osquery (cf. Section~\ref{sec:system}) with an increasing number of osquery hosts and an increasing number of events. For that, we implemented a simplified prototype of our actual Zeek-enhanced osquery implementation in Python that simulates new processes and sockets. Each host continuously sends four events per second to Zeek, which is more then in our real-world evaluation.

\begin{figure}[t]
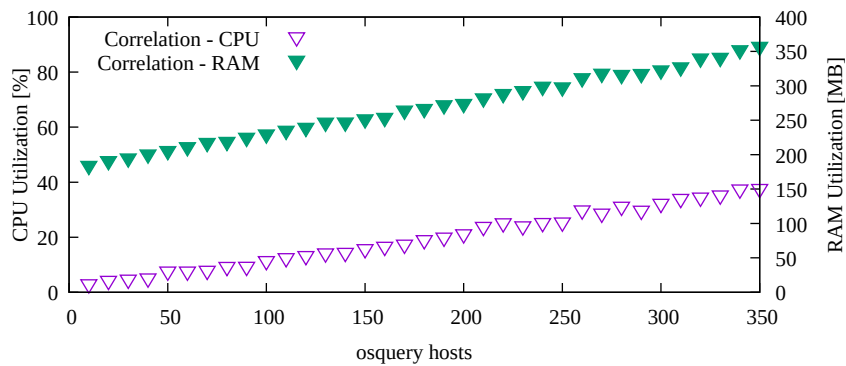

	\centering
	\begin{gnuplot}[terminal=pdf, terminaloptions={color size 12.0cm,5cm}]
		key_name="endpoints"
		hosts="bro"
		changes="2"
		
		set xlabel "osquery hosts"
		set xrange [0:350]
		set yrange [0:100]
		set ylabel "CPU Utilization [
		set y2range [0:]
		set y2label "RAM Utilization [MB]" offset -1
		set ytics nomirror
		set y2tics
		set terminal pdf font "Times New Roman,"
		
		set key left top
		
		plot './'.hosts.'_'.'cpu_max'.'_'.key_name.'_'.'c'.changes.'.data' using 1:3 title "Correlation - CPU" axes x1y1 with points pointtype 10, './'.hosts.'_'.'mem'.'_'.key_name.'_'.'c'.changes.'.data' using 1:3 title "Correlation - RAM" axes x1y2 with points pointtype 11
	\end{gnuplot}
	\caption{Resource utilization of Zeek for hosts with four host events per second.}
	\label{fig:zeek_endpoints_res}
	\vspace*{-12px}
\end{figure}

We distribute the total amount of the lightweight osquery instances equally among ten bare-metal machines. Zeek takes over the role of both {Proxy} and {Authoritative} Zeek and is running on another bare-metal machine. This comes close to a real-world deployment, in which Zeek runs on a single machine and osquery instances are distributed on different machines in the network.
All of the machines are equipped with an Intel(R) Core(TM) i5-2400 CPU @ 3.10GHz and 8GB of RAM.
We run each configuration setup for a specific number of hosts for 20 minutes, measuring the average overhead of Zeek in terms of CPU and RAM utilization for handling osquery hosts and processing host events in Figure~\ref{fig:zeek_endpoints_res}. During the experiment, Zeek retrieves, logs, reconstructs state, and correlates process, socket, and user events from hosts as it is done for the attribution of network flows.
The resources scale linearly with an increasing number of osquery hosts. A single host causes 0.11\% CPU and 0.45MB RAM at the Zeek instance during real-time correlation. Theoretically, to achieve 100\% CPU utilization, about $870$ osquery hosts would be required, each sending four events per second.

\section{Conclusion}
\label{sec:conclusion}

In this paper, we introduced \emph{zeek-osquery} as a novel approach to enrich network intrusion detection with host data. For that, zeek-osquery leverages existing OS instrumentation to collect processes and users and correlates them with network flows.
Our open-source implementation performs efficiently in real-time and in a scalable fashion, as different roles in the platform can be distributed.
Our system gives broader network visibility and attributes network flows to users and applications at large scale. 
Compared to a network-based IDS only, e.g., Zeek, the ratio of attributed network flows to applications increases by orders of magnitudes. Zeek-osquery can attribute more than 96\% of the TCP connections to the originating process and the respective users in real-time. On that basis, we can detect malware that encrypt its communication in protocols such as HTTPS, which usually passes a NIDS without detection.
Future work can include more host data, e.g., opened files, and can develop more correlation algorithms on top of zeek-osquery, e.g., to detect distributed and multi-step attacks.

\bibliographystyle{./splncs04}
\bibliography{./haas20dids}
\end{document}